\documentclass[aps,prl,twocolumn,showpacs]{revtex4}
\usepackage{epsfig}

\begin{document}

\date{\today}
\title{ Ellipsoidal Black Hole -- Black Tori Systems
        in 4D Gravity }
\author{Sergiu I. Vacaru}

\affiliation{Department of Physics, California State University of Fresno,\\
 2345 East San Ramon Ave. M/S 37 Fresno, CA 93740-8031,\ U. S. A.
 \\  and \\
Centro Multidisciplinar de Astrofisica - CENTRA, Departamento de Fisica,\\
Instituto Superior Tecnico, Av. Rovisco Pais 1, Lisboa, 1049-001,
Portugal
 \\ {---} \\ E--mails: sergiuvacaru@venus.nipne.ro,  sergiu$_{-}$vacaru@yahoo.com
}

\begin{abstract}
We construct new classes of exact solutions of the 4D vacuum
Einstein equations which describe  ellipsoidal black holes, black
tori and combined black hole -- black tori configurations. The
solutions can be static or with anisotropic polarizations and
running constants. They are defined by off--diagonal metric ansatz
which may be diagonalized with respect to  anholonomic moving
frames.  We examine physical properties of such anholonomic
gravitational configurations and discuss why the anholonomy may
remove the restriction that horizons must be with spherical
 topology.
\end{abstract}
 \vspace{0.5cm}

\maketitle

\section{Introduction}

Torus configurations of matter around black hole -- neutron star objects are
intensively investigated in modern astrophysics \cite{putten}. One considers
that such tori may radiate gravitational radiation powered by the spin
energy of the black hole in the presence of non--axisymmetries; long
gamma--ray bursts from rapidly spinning black hole--torus systems may
represent hypernovae or black hole--neutron star coalescence. Thus the topic
of constructing of exact vacuum and non--vacuum solutions with non--trivial
topology in the framework of general relativity and extra dimension
gravitational theories becomes of special importance and interest.

In the early 1990s, new solutions with non--sperical black hole horizons
(black tori) were found \cite{lemos} for different states of matter and for
locally anti-de Sitter space times; for a recent review, see \cite{rev}.
Static ellipsoidal black hole, black tori, anisotropic wormhole and Taub NUT
metrics and soltionic solutions of the vacuum and non--vacuum Einstein
equations were constructed in Refs. \cite{v,v2}. Non--trivial topology
configurations (for instance, black rings) are intensively studied in extra
dimension gravity \cite{emp,vth,vtor}.

For four dimensional gravity (4D), it is considered that a number of
classical theorems \cite{israel} impose that a stationary, asymptotically
flat, vacuum black hole solution is completely characterized by its mass and
spin and event horizons of non--spherical topology are forbidden \cite{haw};
see \cite{cen} for further discussion of this issue.

Nevertheless, there were constructed various classes of exact solutions in
4D and 5D gravity with non--trivial topology, anisotropies, solitonic
configurations, running constants and warped factors, under certain
conditions defining static configurations in 4D vacuum gravity. Such metrics
were parametrized by off--diagonal ansazt (for coordinate frames) which can
be effectively diagonalized with respect to certain anholonomic frames with
mixtures of holonomic and anholonomic variables. The system of vacuum
Einstein equations for such ansatz becomes exactly integrable and describe a
new "anholonomic nonlinear dynamics" of vacuum gravitational fields, which
posses generic local anisotropy. The new classes of solutions may have
locally isotropic limits, or can be associated to metric coefficients of
some well known, for instance, black hole, cylindrical, or wormhole soutions.

There is one important question if such anholonomic (anisotropic) solutions
can exist only in extra dimension gravity, with some specific effective
reductions to lower dimensions, or the anholonomic transforms generate a new
class of solutions even in general relativity theory which might be not
restricted by the conditions of Israel--Carter--Robinson uniqueness and
Hawking cosmic cenzorship theorems \cite{israel,haw}?

In the present paper, we explore possible 4D ellipsoidal black hole -- black
torus systems which are defined by generic off--diagonal matrices and
describe anholonomic vacuum gravitational configurations. We present a new
class of exact solutions of 4D vacuum Einstein equations which can be
associated to some exact solutions with ellipsoidal/toroidal horizons and
signularities, and theirs superpositions, being of static configuration, or,
in general, with nonlinear gravitational polarization and running constants.
We also discuss implications of these anisotropic solutions to gravity
theories and ponder possible ways to solve the problem with topologically
non--trivial and deformed horizons.

The organization of this paper is as follows:\ In Sec. II, we consider
ellipsoidal and torus deformations and anistoropic confromal transforms of
the Schwarzschil metric. We introduce an off--diagonal ansatz which can be
diagonalized by anholonomic transforms and compute the non--trivial
components of the vacuum Einstein equations in Sec. III. In Sec. IV, we
construct and analyze three types of exact static solutions with
ellisoidal--torus horizons. Sec. V is devoted to generalization of such
solutions for configurations with running constants and anisotropic
polarizations. The conclusion and discussion are presented in Sec. VI.

\section{Ellipsoidal/Torus Deformations of Metrics}

In this Section we analyze\ anholonomic transforms with ellipsoidal/torus
deformations of the Scwarzschild solution to some off--diagonal metrics. We
define the conditions when the \ new 'deformed' metrics are exact solutions
of vacuum Einstein equations.

The Schwarzschild solution may be written in {\it \ isotropic spherical
coordinates} $(\rho ,\theta ,\varphi )$ \thinspace \cite{ll}
\begin{eqnarray}
dS^{2} &=&-\rho _{g}^{2}\left( \frac{\widehat{\rho }+1}{\widehat{\rho }}%
\right) ^{4}\left( d\widehat{\rho }^{2}+\widehat{\rho }^{2}d\theta ^{2}+%
\widehat{\rho }^{2}\sin ^{2}\theta d\varphi ^{2}\right)  \label{schw} \\
&&+\left( \frac{\widehat{\rho }-1}{\widehat{\rho }+1}\right) ^{2}dt^{2},
\nonumber
\end{eqnarray}
where the isotropic radial coordinate $\rho $ is related with the usual
radial coordinate $r$ via the relation $r=\rho \left( 1+r_{g}/4\rho \right)
^{2}$ for $r_{g}=2G_{[4]}m_{0}/c^{2}$ being the 4D gravitational radius of a
point particle of mass $m_{0},$ $G_{[4]}=1/M_{P[4]}^{2}$ is the 4D Newton
constant expressed via Plank mass $M_{P[4]}.$ In our further considerations,
we put the light speed constant $c=1$ and re--scale the isotropic radial
coordinate as $\widehat{\rho }=\rho /\rho _{g},$ with $\rho _{g}=r_{g}/4.$
The metric (\ref{schw}) is a vacuum static solution of 4D Einstein equations
with spherical symmetry describing the gravitational field of a point
particle of mass $m_{0}.$ It has a singularity for $r=0$ and a spherical
horizon for $r=r_{g},$ or, in re--scaled isotropic coordinates, for $%
\widehat{\rho }=1.$ We emphasize that this solution is parametrized by a
diagonal metric given with respect to holonomic coordinate frames.

We may introduce a new 'exponential' radial coordinate $\varsigma =\ln
\widehat{\rho }$ and write the Schwarzschild metric as
\begin{eqnarray}
ds^{2} &=&-\rho _{g}^{2}b\left( \varsigma \right) \left( d\varsigma
^{2}+d\theta ^{2}+\sin ^{2}\theta d\varphi ^{2}\right) +a\left( \varsigma
\right) dt^{2},  \label{schw1} \\
a\left( \varsigma \right) &=&\left( \frac{\exp \varsigma -1}{\exp \varsigma
+1}\right) ^{2},b\left( \varsigma \right) =\frac{\left( \exp \varsigma
+1\right) ^{4}}{(\exp \varsigma )^{2}}.  \label{scw1c}
\end{eqnarray}
The condition of vanishing of coefficient $a\left( \varsigma \right) ,\exp
\varsigma =1,$ defines the horizon 3D spherical hypersurface
\[
\varsigma =\varsigma \left[ \widehat{\rho }\left( \sqrt{x^{2}+y^{2}+z^{2}}%
\right) \right] ,
\]
where $x,y$ and $z$ are usual Cartezian coordinates.

The 3D spherical line element
\[
ds_{(3)}^{2}=d\varsigma ^{2}+d\theta ^{2}+\sin ^{2}\theta d\varphi ^{2},
\]
may be written in arbitrary ellipsoidal, or toroidal, coordinates which
transforms the spherical horizon equation into very sophysticate relations
(with respect to new coordinates).

Our idea is to deform (renormalize) the coefficients (\ref{scw1c}), $a\left(
\varsigma \right) \rightarrow A\left( \varsigma ,\theta \right) $ and $%
b\left( \varsigma \right) \rightarrow B\left( \varsigma ,\theta \right) ,$
as they would define a rotation ellipsoid and/or a toroidal horizon and
symmetry (for simplicity, we shall consider the ellongated ellipsoid
configuration; the flattened ellipsoids may be analyzed in a similar
manner). \ But such a diagonal metric with respect to ellipsoidal, or
toroidal, local coordinate frame does not solve the vacuum Einstein
equations. In order to generate a new vacuum solution we have to
''elongate'' the differentials $d\varphi $ and $dt,$ i. e. to introduce some
''anholonomic transforms'' (see details in \cite{vth}), like
\begin{eqnarray*}
d\varphi &\rightarrow &\delta \varphi +w_{\varsigma }\left( \varsigma
,\theta ,v\right) d\varsigma +w_{\theta }\left( \varsigma ,\theta ,v\right)
d\theta , \\
dt &\rightarrow &\delta t+n_{\varsigma }\left( \varsigma ,\theta ,v\right)
d\varsigma +n_{\theta }\left( \varsigma ,\theta ,v\right) d\theta ,
\end{eqnarray*}
for $v=\varphi $ (static configuration), or $v=t$ (running in time
configuration) and find the conditions when $w$- and $n$--coefficients and
the renormalized metric coefficients define off--diagonal metrics solving
the Einstein equations and possessing some ellipsoidal and/or toroidal
horizons and symmetries.

We shall define the 3D space ellipsoid -- toroidal configuration in this
manner: in the center of Cartezian coordinates we put an rotation ellipsoid
ellongated along axis $z$ (its intersection by the $xy$--coordinat plane
describes a circle of radius $\rho _{g}^{[e]}=\sqrt{x^{2}+y^{2}}\sim \rho
_{g});$ the ellipsoid is surrounded by a torus with the same $z$ axis of
symmetry, when $-z_{0}\leq z\leq z_{0},$ and the interections of the torus
with the $xy$--coordinate plane describe two circles of radia $\rho
_{g}^{[t]}-z_{0}=\sqrt{x^{2}+y^{2}}$ and $\rho _{g}^{[t]}+z_{0}=\sqrt{%
x^{2}+y^{2}};$ the parameters $\rho _{g}^{[e]},\rho _{g}^{[t]}$ and $z_{0}$
are chosen as to define not intersecting toroidal and ellipsoidal horizons,
i. e. the conditions
\begin{equation}
\rho _{g}^{[t]}-z_{0}>\rho _{g}^{[e]}>0  \label{scale}
\end{equation}
are imposed.

\subsection{Ellipsoidal Configurations}

We shall consider the {\it \ rotation ellipsoid coordinates} \cite{korn} $%
(u,\lambda ,\varphi )$ with $0\leq u<\infty ,0\leq \lambda \leq \pi ,0\leq
\varphi \leq 2\pi ,$ where $\sigma =\cosh u\geq 1,$ are related with the
isotropic 3D Cartezian coordinates $\left( x,y,z\right) $ as
\begin{eqnarray}
(x &=&\widetilde{\rho }\sinh u\sin \lambda \cos \varphi ,  \label{rec} \\
y &=&\widetilde{\rho }\sinh u\sin \lambda \sin \varphi ,z=\widetilde{\rho }%
\cosh u\cos \lambda )  \nonumber
\end{eqnarray}
and define an elongated rotation ellipsoid hypersurface
\begin{equation}
\left( x^{2}+y^{2}\right) /(\sigma ^{2}-1)+\tilde{z}^{2}/\sigma ^{2}=%
\widetilde{\rho }^{2}.  \label{reh}
\end{equation}
with $\sigma =\cosh u.$ The 3D metric on a such hypersurface is
\[
dS_{(3D)}^{2}=g_{uu}du^{2}+g_{\lambda \lambda }d\lambda ^{2}+g_{\varphi
\varphi }d\varphi ^{2},
\]
where
\begin{eqnarray*}
g_{uu} &=&g_{\lambda \lambda }=\widetilde{\rho }^{2}\left( \sinh ^{2}u+\sin
^{2}\lambda \right) , \\
g_{\varphi \varphi } &=&\widetilde{\rho }^{2}\sinh ^{2}u\sin ^{2}\lambda .
\end{eqnarray*}

We can relate the rotation ellipsoid coordinates\newline
$\left( u,\lambda ,\varphi \right) $ from (\ref{rec}) with the isotropic
radial coordinates $\left( \widehat{\rho },\theta ,\varphi \right) $, scaled
by the constant $\rho _{g}, $ from (\ref{schw}), $\ $equivalently with
coordinates $\left( \varsigma ,\theta ,\varphi \right) $ from (\ref{schw1}),
as
\[
\widetilde{\rho }=1,\cosh u=\widehat{\rho }=\exp \varsigma
\]
and deform the Schwarzschild metric by introducing ellipsoidal coordinates
and a new horizon defined by the condition that vanishing of the metric
coefficient before $dt^{2}$ describe an elongated rotation ellipsoid
hypersurface (\ref{reh}),
\begin{eqnarray}
ds_{E}^{2} &=&-\rho _{g}^{2}\left( \frac{\cosh u+1}{\cosh u}\right)
^{4}(\sinh ^{2}u+\sin ^{2}\lambda )  \label{schel} \\
&&\times \lbrack du^{2}+d\lambda ^{2}+\frac{\sinh ^{2}u~\sin ^{2}\lambda }{%
\sinh ^{2}u+\sin ^{2}\lambda }d\varphi ^{2}]  \nonumber \\
& &+\left( \frac{\cosh u-1}{\cosh u+1}\right) ^{2}dt^{2}.  \nonumber
\end{eqnarray}
The ellipsoidally deformed metric (\ref{schel}) does not satisfy the vacuum
Einstein equations, but at long distances from the horizon it transforms
into the usual Schwarzchild solution (\ref{schw}).

We introduce two Classes (A and B) of 4D auxiliary pseudo--Riemannian
metrics, also given in ellipsoid coordinates, being some conformal
transforms of (\ref{schel}), like
\[
ds_{E}^{2}=\Omega _{A(B)E}\left( u,\lambda \right) ds_{A(B)E}^{2}
\]
which also are not supposed to be solutions of the Einstein equations:

Metric of Class A:
\begin{equation}
ds_{(AE)}^{2}=-du^{2}-d\lambda ^{2}+a_{E}(u,\lambda )d\varphi
^{2}+b_{E}(u,\lambda )dt^{2}],  \label{auxm1}
\end{equation}
where
\begin{eqnarray}
a_{E}(u,\lambda ) &=&-\frac{\sinh ^{2}u~\sin ^{2}\lambda }{\sinh ^{2}u+\sin
^{2}\lambda },  \label{auxm1s} \\
b_{E}(u,\lambda ) &=&\frac{(\cosh u-1)^{2}\cosh ^{4}u}{\rho _{g}^{2}(\cosh
u+1)^{6}(\sinh ^{2}u+\sin ^{2}\lambda )},  \nonumber
\end{eqnarray}
which results in the metric (\ref{schel}) by multiplication on the conformal
factor
\begin{equation}
\Omega _{AE}\left( u,\lambda \right) =\rho _{g}^{2}\frac{(\cosh u+1)^{4}}{%
\cosh ^{4}u}(\sinh ^{2}u+\sin ^{2}\lambda ).  \label{auxm1c}
\end{equation}

Metric of Class B:
\begin{equation}
ds_{BE}^{2}=g_{E}(u,\lambda )\left( du^{2}+d\lambda ^{2}\right) -d\varphi
^{2}+f_{E}(u,\lambda )dt^{2},  \label{auxm2}
\end{equation}
where
\begin{eqnarray}
g_{E}(u,\lambda ) &=&-\frac{\sinh ^{2}u+\sin ^{2}\lambda }{\sinh ^{2}u~\sin
^{2}\lambda },  \nonumber \\
f_{E}(u,\lambda ) &=&\frac{(\cosh u-1)^{2}\cosh ^{4}u}{\rho _{g}^{2}(\cosh
u+1)^{6}\sinh ^{2}u\sin ^{2}\lambda },  \label{auxm2f}
\end{eqnarray}
which results in the metric (\ref{schel}) by multiplication on the conformal
factor
\[
\Omega _{BE}\left( u,\lambda \right) =\rho _{g}^{2}\frac{(\cosh u+1)^{4}}{%
\cosh ^{4}u}\sinh ^{2}u\sin ^{2}\lambda .
\]

In Ref. \cite{vth} we proved that there are anholonomic transforms of the
metrics (\ref{schel}), (\ref{auxm1}) and (\ref{auxm2}) which results in
exact ellipsoidal black hole solutions of the vacuum Einstein equations.

\subsection{Toroidal Configurations}

Fixing a scale parameter $\rho _{g}^{[t]}$ which satisfies the conditions (%
\ref{scale}) we define the {\it \ toroidal coordinates} $(\sigma ,\tau
,\varphi )$ (we emphasize that in in this paper we use different letters for
ellipsoidal and toroidal coordinates introduced in Ref. \cite{korn}). These
coordinates run the values $-\pi \leq \sigma <\pi ,0\leq \tau \leq \infty
,0\leq \varphi <2\pi .$ They are related with the isotropic 3D Cartezian
coordinates via transforms
\begin{eqnarray}
\tilde{x} &=&\frac{\widetilde{\rho }\sinh \tau }{\cosh \tau -\cos \sigma }%
\cos \varphi ,  \label{rect} \\
\tilde{y} &=&\frac{\widetilde{\rho }\sinh \tau }{\cosh \tau -\cos \sigma }%
\sin \varphi ,\tilde{z}=\frac{\widetilde{\rho }\sinh \sigma }{\cosh \tau
-\cos \sigma }  \nonumber
\end{eqnarray}
and define a toroidal hypersurface
\[
\left( \sqrt{\tilde{x}^{2}+\tilde{y}^{2}}-\widetilde{\rho }\frac{\cosh \tau
}{\sinh \tau }\right) ^{2}+\tilde{z}^{2}=\frac{\widetilde{\rho }^{2}}{\sinh
^{2}\tau }.
\]
The 3D metric on a such toroidal hypersurface is
\[
ds_{(3D)}^{2}=g_{\sigma \sigma }d\sigma ^{2}+g_{\tau \tau }d\tau
^{2}+g_{\varphi \varphi }d\varphi ^{2},
\]
where
\begin{eqnarray*}
g_{\sigma \sigma } &=&g_{\tau \tau }=\frac{\widetilde{\rho }^{2}}{\left(
\cosh \tau -\cos \sigma \right) ^{2}}, \\
g_{\varphi \varphi } &=&\frac{\widetilde{\rho }^{2}\sinh ^{2}\tau }{\left(
\cosh \tau -\cos \sigma \right) ^{2}}.
\end{eqnarray*}

We can relate the toroidal coordinates $\left( \sigma ,\tau ,\varphi \right)
$ from (\ref{rect}) with the isotropic radial coordinates $\left( \widehat{%
\rho }^{[t]},\theta ,\varphi \right) $, scaled by the constant $\rho
_{g}^{[t]},$ as
\[
\widetilde{\rho }=1,\sinh ^{-1}\tau =\widehat{\rho }^{[t]}
\]
and transform the Schwarzschild solution into a new metric with toroidal
coordinates \ by changing the 3D radial line element into the toroidal one
and stating the $tt$--coefficient of the metric to have a toroidal horizon.
The resulting metric is
\begin{eqnarray}
ds_{T}^{2} &=&-\left( \rho _{g}^{[t]}\right) ^{2}\frac{\left( \sinh \tau
+1\right) ^{4}}{\left( \cosh \tau -\cos \sigma \right) ^{2}} \times
\label{schtor} \\
&& \left( d\sigma ^{2}+d\tau ^{2}+\sinh ^{2}\tau d\varphi ^{2}\right) +
\left( \frac{\sinh \tau -1}{\sinh \tau +1}\right) ^{2}dt^{2},  \nonumber
\end{eqnarray}
Such a deformed Schwarzchild like toroidal metric is not an exact solution
of the vacuum Einstein equations, but at long radial distances it transforms
into the usual Schwarzchild solution with effective horizon $\rho _{g}^{[t]}$
with the 3D line element parametrized by toroidal coordinates.

We introduce two Classes (A and B) of 4D auxiliary pseudo-Riemannian
metrics, also given in toroidal coordinates, being some conformal transforms
of (\ref{schtor}), like
\[
ds_{T}^{2}=\Omega _{A(B)T}\left( \sigma ,\tau \right) ds_{A(B)T}^{2}
\]
but which are not supposed to be solutions of the Einstein equations:

Metric of Class A:
\begin{equation}
ds_{AT}^{2}=-d\sigma ^{2}-d\tau ^{2}+a_{T}(\tau )d\varphi ^{2}+b_{T}(\sigma
,\tau )dt^{2},  \label{auxm1t}
\end{equation}
where
\begin{eqnarray}
a_{T}(\tau ) &=&-\sinh ^{2}\tau ,  \nonumber \\
b_{T}(\sigma ,\tau ) &=&\frac{\left( \sinh \tau -1\right) ^{2}\left( \cosh
\tau -\cos \sigma \right) ^{2}}{\rho _{g}^{[t]2}\left( \sinh \tau +1\right)
^{6}},  \label{auxm1tb}
\end{eqnarray}
which results in the metric (\ref{schtor}) by multiplication on the
conformal factor
\begin{equation}
\Omega _{AT}\left( \sigma ,\tau \right) =\rho _{g}^{[t]2}\frac{\left( \sinh
\tau +1\right) ^{4}}{\left( \cosh \tau -\cos \sigma \right) ^{2}}.
\label{auxm1tc}
\end{equation}

Metric of Class B:
\begin{equation}
ds_{BT}^{2}=g_{T}(\tau )\left( d\sigma ^{2}+d\tau ^{2}\right) -d\varphi
^{2}+f_{T}(\sigma ,\tau )dt^{2},  \label{auxm2t}
\end{equation}
where
\begin{eqnarray*}
g_{T}(\tau ) &=&-\sinh ^{-2}\tau , \\
f_{T}(\sigma ,\tau ) &=&\rho _{g}^{[t]2}\left( \frac{\sinh ^{2}\tau -1}{%
\cosh \tau -\cos \sigma }\right) ^{2},
\end{eqnarray*}
which results in the metric (\ref{schtor}) by multiplication on the
conformal factor
\begin{equation}
\Omega _{BT}\left( \sigma ,\tau \right) =\left( \rho _{g}^{[t]}\right) ^{-2}%
\frac{\left( \cosh \tau -\cos \sigma \right) ^{2}}{\left( \sinh \tau
+1\right) ^{2}}.  \label{auxm2tbc}
\end{equation}

In Ref. \cite{vtor} we used the metrics (\ref{schtor}), (\ref{auxm1t}) and (%
\ref{auxm2t}) in order to generate exact solutions of the Einstein equations
with toroidal horizons and anisotropic polarizations and running constants
by performing corresponding anholonomic transforms.

\section{The Metric Ansatz and Vacuum Einstein Equations}

Let us denote the local system of coordinates as $u^{\alpha }=\left(
x^{i},y^{a}\right) ,$ where $\ x^{1}=u$ and $x^{2}=\lambda $ for ellipsoidal
coordinates ($x^{1}=\sigma $ and $x^{2}=\tau $ for toroidal coordinates) and
$y^{3}=v=\varphi $ and $y^{4}=t$ for the so--called $\varphi $--anisotropic
configurations ($y^{4}=v=t$ and $y^{5}=\varphi $ for the so--called $t$%
--anisotropic configurations). Our spacetime is modeled as a 4D
pseudo--Riemannian space of signature $\left( -,-,-,+\right) $ (or $\left(
-,-,+,-\right) ),$ which in general may be enabled with an anholonomic frame
structure (tetrads, or vierbiend) $e_{\alpha }=A_{\alpha }^{\beta }\left(
u^{\gamma }\right) \partial /\partial u^{\beta }$ subjected to some
anholonomy \ relations
\begin{equation}
e_{\alpha }e_{\beta }-e_{\beta }e_{\alpha }=W_{\alpha \beta }^{\gamma
}\left( u^{\varepsilon }\right) e_{\gamma },  \label{anhol}
\end{equation}
where $W_{\alpha \beta }^{\gamma }\left( u^{\varepsilon }\right) $ are
called the coefficients of anholonomy.

The anholonomically and conformally transformed 4D line element is
\begin{equation}
ds^{2}=\Omega ^{2}(x^{i},v)\hat{{g}}_{\alpha \beta }\left( x^{i},v\right)
du^{\alpha }du^{\beta },  \label{cmetric4}
\end{equation}
were the coefficients $\hat{{g}}_{\alpha \beta }$ are parametrized by the
ansatz {\scriptsize
\begin{equation}
\left[
\begin{array}{cccc}
g_{1}+\zeta _{1}^{\ 2}h_{3}+n_{3}^{\ 2}h_{4} & \zeta _{1}\zeta
_{2}h_{3}+n_{1}n_{2}h_{4} & \zeta _{1}h_{3} & n_{1}h_{4} \\
\zeta _{1}\zeta _{2}h_{3}+n_{1}n_{2}h_{4} & g_{2}+\zeta _{2}^{\
2}h_{3}+n_{3}^{\ 2}h_{4} & +\zeta _{2}h_{3} & n_{2}h_{4} \\
\zeta _{1}h_{3} & \zeta _{2}h_{3} & h_{3} & 0 \\
n_{1}h_{4} & n_{2}h_{4} & 0 & h_{4}
\end{array}
\right],  \label{ansatzc4}
\end{equation}
} with $g_{i}=g_{i}\left( x^{i}\right) ,h_{a}=h_{ai}\left( x^{k},v\right)
,n_{i}=n_{i}\left( x^{k},v\right) ,$ $\zeta _{i}=\zeta _{i}\left(
x^{k},v\right) $ $\ $\ and $\Omega =\Omega \left( x^{k},v\right) $ being
some functions of necessary smoothly class or even sigular in some points
and finite regions. So, the $g_{i}$--components of our ansatz depend only on
''holonomic'' variables $x^{i}$ and the rest of coefficients may also depend
on ''anisotropic'' (anholonomic) variable $y^{3}=v;$ our ansatz does not
depend on the second anisotropic variable $y^{4}.$

We may diagonalize the line element
\begin{equation}
\delta s^{2}=\Omega ^{2}[g_{1}(dx^{1})^{2}+g_{2}(dx^{2})^{2}+h_{3}(\delta
v)^{2}+h_{4}(\delta y^{4})^{2}],  \label{dmetric4}
\end{equation}
with respect to the anholonomic co--frame \newline
$\delta ^{\alpha }=\left( dx^{i},\delta v,\delta y^{4}\right) ,$ where
\begin{equation}
\delta v=dv+\zeta _{i}dx^{i}\mbox{ and }\delta y^{4}=dy^{4}+n_{i}dx^{i},
\label{ddif4}
\end{equation}
which is dual to the frame $\delta _{\alpha }=\left( \delta _{i},\partial
_{4},\partial _{5}\right) ,$ where
\begin{equation}
\delta _{i}=\partial _{i}+\zeta _{i}\partial _{3}+n_{i}\partial _{4}.
\label{dder4}
\end{equation}
The tetrads $\delta _{\alpha }$ and $\delta ^{\alpha }$ are anholonomic
because, in general, they satisfy some non--trivial anholonomy relations (%
\ref{anhol}). The anholonomy is induced by the coefficients $\zeta _{i}$ and
$n_{i}$ which ''elongate'' partial derivatives and differentials if we are
working with respect to anholonomic frames. This result in a more
sophisticate differential and integral calculus (a usual situation in
'tetradic' and 'spinor' gravity), but simplifies substantially tensor
computations, because we are dealing with diagonalized metrics.

The vacuum Einstein equations for the (\ref{ansatzc4}) (equivalently, for (%
\ref{dmetric4})), $R_{\alpha }^{\beta }=0,$ computed with respect to
anholonomic frames (\ref{ddif4}) and (\ref{dder4}), transforms into a system
of partial differential equations \cite{v,vth,vtor}:
\begin{eqnarray}
R_{1}^{1}=R_{2}^{2}=-\frac{1}{2g_{1}g_{2}}[g_{2}^{\bullet \bullet }-\frac{%
g_{1}^{\bullet }g_{2}^{\bullet }}{2g_{1}}-\frac{(g_{2}^{\bullet })^{2}}{%
2g_{2}} &&  \nonumber \\
+g_{1}^{^{\prime \prime }}-\frac{g_{1}^{^{\prime }}g_{2}^{^{\prime }}}{2g_{2}%
}-\frac{(g_{1}^{^{\prime }})^{2}}{2g_{1}}] &=&0,  \label{ricci1a} \\
R_{3}^{3}=R_{4}^{4}=\frac{-1}{2h_{3}h_{4}}\left[ h_{4}^{\ast \ast
}-h_{4}^{\ast }\left( \ln \sqrt{|h_{3}h_{4}|}\right) ^{\ast }\right] &=&0,
\label{ricci2a} \\
R_{4i}=-\frac{h_{4}}{2h_{3}}\left[ n_{i}^{\ast \ast }+\gamma n_{i}^{\ast }%
\right] &=&0,  \label{ricci4a}
\end{eqnarray}
where
\begin{equation}
\gamma =3h_{4}^{\ast }/2h_{4}-h_{3}^{\ast }/h_{3},  \label{abc}
\end{equation}
and the partial derivatives are abreviated like $g_{1}^{\bullet }=\partial
g_{1}/\partial x^{1},g_{1}^{^{\prime }}=\partial g_{1}/\partial x^{2}$ and $%
h_{3}^{\ast }=\partial h_{3}/\partial v.$ The coefficients $\zeta _{{i}}$
are found as to consider non--trivial conformal factors $\Omega :$ we
compensate by $\zeta _{{i}}$ possible conformal deformations of the Ricci
tensors, computed with respect to anholonomic frames. The conformal
invariance of such anholonomic transforms holds if
\begin{equation}
\Omega ^{q_{1}/q_{2}}=h_{3}~(q_{1}\mbox{ and }q_{2}\mbox{ are
integers}),  \label{confq}
\end{equation}
and there are satisfied the equations
\begin{equation}
\partial _{i}\Omega -\zeta _{{i}}\Omega ^{\ast }=0.  \label{confeq}
\end{equation}

The system of equations (\ref{ricci1a})--(\ref{ricci4a}) and (\ref{confeq})
can be integrated in general form \cite{vth}. Physical solutions are defined
from some additional boundary conditions, imposed types of symmetries,
nonlinearities and singular behaviour and compatibility in locally
anisotropic limits with some well known exact solutions.

In this paper we give some examples of ellipsoidal and toroidal solutions
and investigate some classes of metrics for combined ellipsoidal black hole
-- black tori configurations.

\section{Static Black Hole -- Black Torus Metrics}

We analyzed in detail the method of anholonomic frames and constructed 4D
and 5D ellipsoidal black hole and black tori solutions in Refs. \cite
{v,vth,vtor}. In this Section we give same new examples of metrics
describing \ one static 4D black hole \ or one static 4D black torus
configurations. Then we extend the constructions for metrics describing
combined variants of black hole -- black torus solutions. We shall analyze
solutions with trivial and non--trivial conformal factors.

In this section the 4D local coordinates are written as $\left(
x^{1},x^{2},y^{3}=v=\varphi ,y^{4}=t\right) ,$ where we take $%
x^{i}=(u,\lambda )$ for ellipsoidal configurations and $x^{i}=(\sigma ,\tau
) $ for toroidal configurations. Here we note that, we can introduce a
''general'' 2D space ellipsoidal coordinate system, $u=u(\sigma ,\tau )$ and
$\lambda =\tau ,$\ for both ellipsoidal and toroidal configurations if, for
instance, we identify the ellipsoidal coordinate $\lambda $ with the
toroidal $\tau ,$ and relate $u$ with $\sigma $ and $\tau $ as
\[
\sinh u=\frac{1}{\cosh \tau -\cos \sigma }.
\]
In the vicinity of $\tau =0$ we can aproximate $\cosh \tau \approx 1$ and to
write $u=u\left( \sigma \right) $and $\lambda =\tau .$ For $\tau \gg 1$ we
have
\[
\sinh u\approx \frac{1}{\cosh \tau }\left( 1+\frac{1}{\cos \sigma }\right) .
\]
In general, we consider that the ''holonomic'' coordinates are some
functions $x^{i}=x^{i}\left( \sigma ,\tau \right) =\widetilde{x}^{i}\left(
u,\lambda \right) $ for which the 2D line element can be written in
conformal metric form,
\[
ds_{[2]}^{2}=-\mu ^{2}\left( x^{i}\right) \left[ \left( dx^{1}\right)
^{2}+\left( dx^{2}\right) ^{2}\right] .
\]
For simlicity, we consider 4D coordinate parametrizations when the angular
coordinate $\varphi $ and the time like coordinate $t$ are not affected by
any transforms of $x$-coordinates.

\subsection{Static anisotropic black hole/torus solutions}

\subsubsection{An example of ellipsoidal\ black hole configuration}

The symplest way to generate a static but anisotropic ellipsoidal black hole
solution with an anholonomically diagonalized metric (\ref{dmetric4}) is \
to take a metric of type (\ref{auxm1}), to ''elongate'' its differentials,
\begin{eqnarray*}
d\varphi &\rightarrow &\delta \varphi =d\varphi +\zeta _{i}\left(
x^{k},\varphi \right) dx^{i}, \\
dt &\rightarrow &\delta t=dt+n_{i}\left( x^{k},\varphi \right) dx^{i},
\end{eqnarray*}
than to multiply on a conformal factor
\[
\Omega ^{2}\left( x^{k},\varphi \right) =\omega ^{2}\left( x^{k},\varphi
\right) \Omega _{AE}^{2}(x^{k}),
\]
the factor $\omega ^{2}\left( x^{k},\varphi \right) $ is obtained by
rescaling the consant $\rho _{g}$ from (\ref{auxm1c}),
\begin{equation}
\rho _{g}\rightarrow \overline{\rho }_{g}=\omega \left( x^{k},\varphi
\right) \rho _{g},  \label{oresc}
\end{equation}
in the simplest case we can consider only ''angular'' on $\varphi $
anisotropies. Then we 'renormalize' (by introducing $x^{i}$ coordinates) the
$g_{1},g_{2}$ and $h_{3}$ coefficients,
\begin{eqnarray}
g_{1,2} &=&-1\rightarrow -\mu ^{2}\left( x^{i}\right) ,  \label{parsol1} \\
h_{3} &=&h_{3[0]}=a_{E}(u,\lambda )\rightarrow h_{3}=-\Omega ^{-2}\left(
x^{k},\varphi \right) ,  \label{parsol1a}
\end{eqnarray}
we fix a relation of type (\ref{confq}), and take $h_{4}=$ $%
h_{4[0]}=b_{E}(x^{i}).$ The anholonomically transformed metric is
pa\-ra\-metrized in the form
\begin{eqnarray}
\delta s^{2} &=&\Omega ^{2}\{-\mu ^{2}\left( x^{i}\right) \left[
(dx^{1})^{2}+(dx^{2})^{2}\right]  \label{sol1} \\
&&-\Omega ^{-2}\left( x^{k},\varphi \right) (\delta
v)^{2}+b_{E}(x^{i})(\delta y^{4})^{2}\},  \nonumber
\end{eqnarray}
where $\mu ,\zeta _{i}$ and $n_{i}$ are to be defined respectively from the
equations (\ref{ricci1a}), (\ref{confeq}) and (\ref{ricci4a}). We note that
the equation (\ref{ricci2a}) is already solved because in our case $%
h_{4}^{\ast }=0.$

The equation (\ref{ricci1a}), with partial derivations on coordinates $x^{i}
$ and parametrizations (\ref{parsol1}) has the general solution
\begin{equation}
\mu ^{2}=\mu _{\lbrack 0]}^{2}\exp \left[ c_{[1]}x^{1}\left( u,\lambda
\right) +c_{[2]}x^{2}\left( u,\lambda \right) \right] ,  \label{f1}
\end{equation}
where $\mu _{\lbrack 0]},c_{[1]}$ and $c_{[2]}$ are some constants which
should be defined from boundary conditions and by fixing a corresponding 2D
system of coordinates; we pointed that we may redefine the factor (\ref{f1})
in 'pure' ellipsoidal coordinates $\left( u,\lambda \right) .$

The general solution of (\ref{confeq}) for renormalization (\ref{oresc}) and
parametrization (\ref{parsol1a}) is
\begin{eqnarray}
\zeta _{i}\left( x^{k},\varphi \right) &=&\left( \omega ^{\ast }\right)
^{-1}\partial _{i}\omega +\partial _{i}\ln |\Omega _{AE}|/\left( \ln |\omega
|\right) ^{\ast },  \label{e1} \\
&=&\partial _{i}\ln |\Omega _{AE}|/\left( \ln |\omega |\right) ^{\ast }%
\mbox{ for }\omega =\omega \left( \varphi \right) .  \nonumber
\end{eqnarray}
For a given $h_{3}$ with $h_{4}^{\ast }=0,$ we can compute the coefficient $%
\gamma $ from (\ref{abc}). After two integrations on $\varphi $ in (\ref
{ricci4a}) we find
\begin{equation}
n_{i}\left( x^{k},\varphi \right) =n_{i[0]}\left( x^{k}\right)
+n_{i[1]}\left( x^{k}\right) \int \omega ^{-2}d\varphi .  \label{n1}
\end{equation}
The set of functions (\ref{f1}), (\ref{e1}) and (\ref{n1}) for any given $%
\Omega _{AE}\left( x^{i}\right) $ and $\omega \left( x^{k},\varphi \right) $
defines an exact static solution of the vacuum Einstein equations
parametrized by an off--diagonal metric of type (\ref{sol1}). This solution
have an ellipsoidal horizon defined by the condition of vanishing of the
coefficient $h_{4[0]}=b_{E}(x^{i}),$ see the coefficients for the auxiliarry
metric (\ref{auxm1}) and an anisotropic effective constant (\ref{oresc}).
This is a general solution depending on arbitrary functions $\omega \left(
x^{k},\varphi \right) $ and $n_{i[0,1]}\left( x^{k}\right) $ and constants $%
\mu _{\lbrack 0]},c_{[1]}$ and $c_{[2]}$ which have to be stated
from some additional physical arguments.

For instance, if we wont to impose the condition that our solution, far away
from the ellipsoidal horizon, transform into the Scwarzshild solution with
an effective anisotropic ''mass'', or a renromalized gravitational Newton
constant, we may put $\mu _{\lbrack 0]}=1$ and fix the $x^{i}$--coordinates
and constants $c_{[1,2]}$ as to obtain the linear interval
\[
ds_{[2]}^{2}=-\left[ du^{2}+d\lambda ^{2}\right] .
\]
The coefficients $n_{i[0,1]}\left( x^{k}\right) $ and $\omega \left(
x^{k},\varphi \right) $ may be taken as at long distances from the horizon
one holds the limits $n_{i[0,1]}\left( x^{k}\right) \rightarrow 0$ and $%
\zeta _{i}\left( x^{k},\varphi \right) \rightarrow 0$ for $\omega \left(
x^{k},\varphi \right) \rightarrow 0.$ In this case, at asymptotics, our
solution will transform into a Schwarzshild like solution with
''renormalized'' parameter $\overline{\rho }_{g}\rightarrow const.$

Nevertheless, we consider that it is not obligatory to select only such type
of ellipsoidal solutions (with imposed asymptotic spherical symmetry)
parametrized by metrics of class (\ref{sol1}). The system of vacuum
gravitational equations (\ref{ricci1a})--(\ref{confeq}) for the ansatz (\ref
{sol1}) defines a nonlinear static configuration (an alternative vacuum
Einstein configuration with ellipsoidal horizon) which, in general, is not
equivalent to the Schwarschild vacuum. This points to some specific
properties of the gravitational vacuum which follow from the nonlinear
character of the Einstein equations. In quantum field theory the nonlinear
effects may result in unitary non--equivalent vacua; in classical
gravitational theories we could obtain a similar behaviour if we are dealing
with off--diagonal metrics and anholonomic frames.

\begin{figure*}
\includegraphics[scale=0.8]{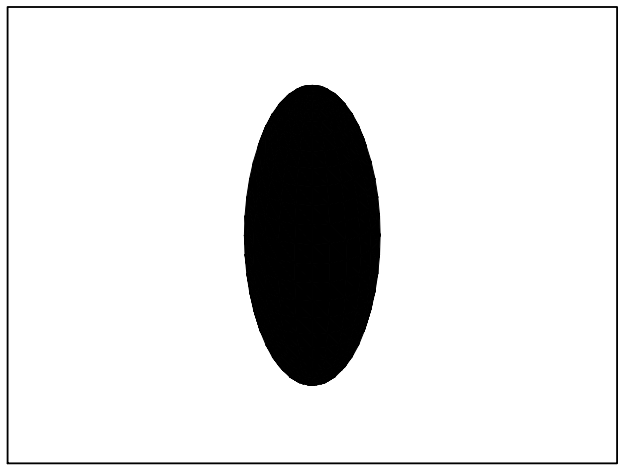}
\caption{\label{ellipsoid}Ellipsoidal Configuration}
\end{figure*}

The constructed new static vacuum solution (\ref{sol1}) for a 4D ellipsoidal
black hole is stated by the coefficients
\begin{eqnarray}
g_{1,2} &=&-1,\mu =1,\overline{\rho }_{g}=\omega \left( x^{k},\varphi
\right) \rho _{g},\Omega ^{2}=\omega ^{2}\Omega _{AE}^{2},  \nonumber \\
h_{3} &=&-\Omega ^{-2}\left( x^{k},\varphi \right) ,h_{4}=b_{E}(x^{i}), (%
\mbox{ see (\ref{auxm1}),(\ref{auxm1c})}),  \nonumber \\
\zeta _{i} &=&\left( \omega ^{\ast }\right) ^{-1}\partial _{i}\omega
+\partial _{i}\ln |\Omega _{AE}|/\left( \ln |\omega |\right) ^{\ast },
\nonumber \\
n_{i} &=&n_{i[0]}\left( x^{k}\right) +n_{i[1]}\left( x^{k}\right) \int
\omega ^{-2}d\varphi .  \label{data1}
\end{eqnarray}
These data define an ellipsoidal configuration, see Fig.
\ref{ellipsoid}.

Finally, we remark that we have generated a vacuum ellipsoidal gravitational
configuration starting from the metric (\ref{auxm1}), i. e. we constructed
an ellipsoidal $\varphi $--solution of Class A (see details on
classification in \cite{vth}). In a similar manner we can define anholonomic
deformations of the metric (\ref{auxm2}) and renormalization of conformal
factor $\Omega _{BE}\left( u,\lambda \right) $ in order to construct an
ellipsoidal $\varphi $--solution of Class B. We omit such considerations in
this paper but present, in the next subsection, an example of toroidal $%
\varphi $--solution of Class B.

\subsubsection{An example of toroidal\ black hole configuration}

We start with the metric (\ref{auxm2t}), ''elongate'' its differentials $%
d\varphi \rightarrow \delta \varphi $ and $dt\rightarrow \delta t$ and than
multiply on a conformal factor
\[
\Omega ^{2}\left( x^{k},\varphi \right) =\varpi ^{2}\left( x^{k},\varphi
\right) \Omega _{BT}^{2}(x^{k})g_{T}\left( \tau \right) ,
\]
see (\ref{auxm2tbc}) which is connected with the renormalization of constant
$\rho _{g}^{[t]},$
\begin{equation}
\rho _{g}^{[t]}\rightarrow \overline{\rho }_{g}^{[t]}=\varpi \left(
x^{k},\varphi \right) \rho _{g}^{[t]}.  \label{osect}
\end{equation}
For toroidal configurations it is naturally to use 2D toroidal holonomic
coordinates $x^{i}=(\sigma ,\tau ).$

\begin{figure*}
\includegraphics[scale=0.8]{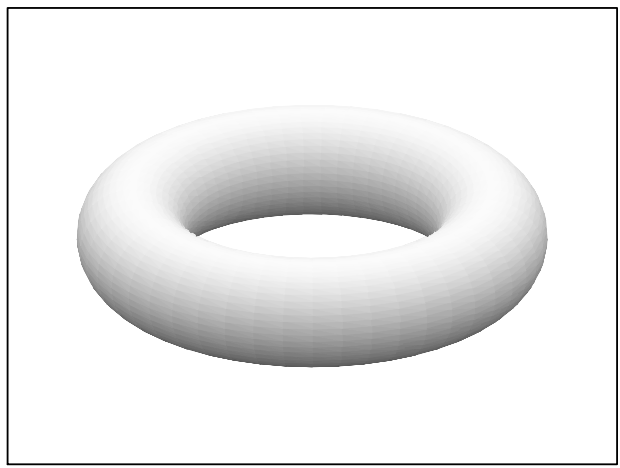}
\caption{\label{torus}Toroidal Configuration}
\end{figure*}

The anholonomically transformed metric is pa\-ra\-met\-rized in the form
\begin{eqnarray}
\delta s^{2} &=&\Omega ^{2}\{-\left[ d\sigma ^{2}+d\tau ^{2}\right] -\eta
_{3}\left( \sigma ,\tau ,\varphi \right) g_{T}^{-1}\left( \tau \right)
\delta \varphi ^{2}  \nonumber \\
&&+f_{T}(\sigma ,\tau )g_{T}^{-1}\left( \tau \right) \delta t^{2}\}.
\label{sol2}
\end{eqnarray}
We state the coefficients
\[
h_{3}=-\eta _{3}\left( \sigma ,\tau ,\varphi \right) g_{T}^{-1}\left( \tau
\right) \mbox{ and }h_{4}=f_{T}(\sigma ,\tau )g_{T}^{-1}\left( \tau \right)
,
\]
where the polarization
\[
\eta _{3}\left( \sigma ,\tau ,\varphi \right) =\varpi ^{-2}\left( \sigma
,\tau ,\varphi \right) \Omega _{BT}^{-2}(\sigma ,\tau )
\]
is found from the condition (\ref{confq}) as $h_{3}=-\Omega ^{-2}.$ The
equation (\ref{ricci2a}) is solved by arbitrary couples $h_{3}\left( \sigma
,\tau ,\varphi \right) $ $\ $and $h_{4}(\sigma ,\tau )$ when $h_{4}^{\ast
}=0.$ The procedure of definition of $\zeta _{i}\left( \sigma ,\tau ,\varphi
\right) $ and $n_{i}\left( \sigma ,\tau ,\varphi \right) $ is similar to
that from the previous subsection. We present the final results as the data
\begin{eqnarray}
g_{1,2} &=&-1,\overline{\rho }_{g}=\varpi \left( \sigma ,\tau ,\varphi
\right) \rho _{g},\Omega ^{2}=\varpi ^{2}\Omega _{BT}^{2}g_{T}\left( \tau
\right) ,  \nonumber \\
h_{3} &=&-\eta _{3}\left( \sigma ,\tau ,\varphi \right) g_{T}^{-1}\left(
\tau \right) ,h_{4}=f_{T}(\sigma ,\tau )g_{T}^{-1}\left( \tau \right) ,
\nonumber \\
\eta _{3} &=&\varpi ^{-2}\left( \sigma ,\tau ,\varphi \right) \Omega
_{BT}^{-2}(\sigma ,\tau ), (\mbox{ see (\ref{auxm2t}),
(\ref{auxm2tbc})}),  \nonumber \\
\zeta _{i} &=&\left( \varpi ^{\ast }\right) ^{-1}\partial _{i}\varpi
+\partial _{i}\ln |\Omega _{BT}\sqrt{g_{T}}|/\left( \ln |\varpi |\right)
^{\ast },  \label{data2} \\
n_{i} &=&n_{i[0]}\left( \sigma ,\tau \right) +n_{i[1]}\left( \sigma ,\tau
\right) \int \varpi ^{-2}d\varphi  \nonumber
\end{eqnarray}
for the ansatz (\ref{sol2}) which defines an exact static
solution of the vacuum Einstein equations with toroidal symmetry,
of Class B, with anisotropic dependence on coordinate $\varphi ,$
see the torus configuration from Fig.  \ref{torus}. The
off--diagonal solution is non--trivial for anisotropic linear
distributions of mass on the circle contained in the torus ring,
or alternatively, if there is a renormalized gravitational
constant with anisotropic dependence on angle $\varphi .$ This
class of solutions have a toroidal horizon defined by the
condition of vanishing of the coefficient $h_{4}$ which holds if
$f_{T}(\sigma ,\tau )=0.$
The functions $\varpi \left( \sigma ,\tau ,\varphi \right) $ and $%
n_{i[0,1]}\left( \sigma ,\tau \right) $ may be stated in a form that at long
distancies from the toroidal horizon the (\ref{sol2}) with data (\ref{data2}%
) will have asymptotics like the Scwarzschild metric. We can also consider
alternative toroidal vacuum configurations. We note that instead of
relations like $h_{3}=-\Omega ^{-2}$ we can use every type $h_{3}\sim \Omega
^{p/q},$ like is stated by (\ref{confq}); it depends on what type of
nonlinear configuration and asymptotic limits we wont to obtain.

We remark also that in a symilar manner we can generate toroidal
configurations of Class A, starting from the auxiliary metric (\ref{auxm1t}%
). In the next subsection we elucidate this possibility by interfering it
with a Class B ellipsoidal configuration.

\subsection{Static Ellipsoidal Black Hole -- Black Torus solutions}

There are different possibilities to combine static ellipsoidal black hole
and black torus solutions as they will give configurations with two
horizons. In this subsection we analyze two such variants. We consider a 2D
system of holonomic coordinates $x^{i},$ which may be used both on the
'ellipsoidal' and 'toroidal' sectors via transforms like $u=u(x^{i}),\lambda
=\tau \left( x^{i}\right) $ and $\sigma =\sigma \left( x^{i}\right) .$

\subsubsection{Ellipsoidal--torus black configurations of Class BA}

We construct a 4D vacuum metric with posses two type of horizons,
ellipsoidal and toroidal one, having both type characteristics like a metic
of Class B for ellipsoidal configurations and a metric of Class A for
toroidal configurations (we conventionally call this ellipsoidal torus
metric to be of Class BA). \ The ansatz is taken
\begin{eqnarray}
\delta s^{2} &=&\Omega ^{2}\{-\mu ^{2}\left( x^{i}\right) \left[
(dx^{1})^{2}+(dx^{2})^{2}\right]  \label{sol3} \\
&&-\eta _{3}\left( x^{k},\varphi \right) a_{T}\left( x^{i}\right) \delta
\varphi ^{2}+\frac{b_{T}(x^{i})f_{E}(x^{i})}{g_{E}(x^{i})}\delta t^{2}\},
\nonumber
\end{eqnarray}
with
\begin{eqnarray*}
\Omega ^{2} &=&\omega ^{2}\left( x^{k},\varphi \right) \varpi ^{2}\left(
x^{k},\varphi \right) \Omega _{AT}^{2}\left( x^{i}\right) \Omega
_{BE}^{2}\left( x^{i}\right) , \\
\eta _{3} &=&-a_{T}^{-1}\left( x^{i}\right) \Omega ^{-2},h_{3}=-\eta
_{3}\left( x^{k},\varphi \right) a_{T}\left( x^{i}\right) , \\
h_{4} &=&b_{T}(x^{i})f_{E}(x^{i})/g_{E}(x^{i}), \\
\mu ^{2} &=&\mu _{\lbrack 0]}^{2}\exp \left[ c_{[1]}x^{1}+c_{[2]}x^{2}\right]
.
\end{eqnarray*}
So, in general we may having both type of anisotropic renormalizations of
constants $\rho _{g}$ and $\rho _{g}^{[t]}$ as in (\ref{oresc}) and (\ref
{osect}). The prolongations of differentials $\delta \varphi $ and $\delta t$
are defined by the\ coefficients

\begin{eqnarray*}
\zeta _{i}\left( x^{k},\varphi \right) &=&\left( \Omega ^{\ast }\right)
^{-1}\partial _{i}\Omega , \\
n_{i}\left( x^{k},\varphi \right) &=&n_{i[0]}\left( x^{k}\right)
+n_{i[1]}\left( x^{k}\right) \int \omega ^{-2}\varpi ^{-2}d\varphi .
\end{eqnarray*}
The constants $\mu _{\lbrack 0]}^{2},c_{[1,2]},$ functions $\omega
^{2}\left( x^{k},\varphi \right) ,\varpi ^{2}\left( x^{k},\varphi \right) $
and $n_{i[0,1]}\left( x^{k}\right) $ \ and relation $h_{3}\sim \Omega ^{p/q}$%
\ may be selected as to obtain at asymptotics a Schwarschild like behaviour.
The metric (\ref{sol3}) has two horizons, a toroidal one, defined by the
condition $b_{T}(x^{i})=0,$ and an ellipsoidal one, defined by the condition
$f_{E}(x^{i})=0$ (see respectively these functions in (\ref{auxm1tb}) \ and (%
\ref{auxm2f})).

\begin{figure*}
\includegraphics[scale=0.8]{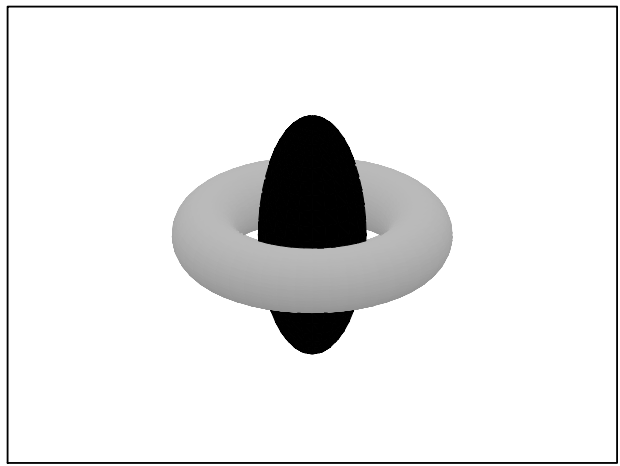}
\caption{\label{eltorus}Ellipsoidal--Torus Configuration}
\end{figure*}

The ellipsoidal--torus configuration is illustrated in  Fig.
\ref{eltorus}.

We can consider different combinations of ellipsoidal black hole an black
torus metrics in order to construct solutions of Class AA, AB and BB (we
omit such similar constructions).

\subsubsection{A second example of ellipsoidal black hole -- black torus
system}

In the simplest case we can construct a solution with an ellipsoidal and
toroidal horizon which have a trivial conformal factor $\Omega $ and
vanishing coefficients $\zeta _{i}=0$ (see (\ref{confeq})). Establishing a
global 3D toroidal space coordinate system, we consider the ansatz
\begin{eqnarray}
\delta s^{2} &=&\{-\left[ d\sigma ^{2}+d\tau ^{2}\right] -\eta _{3}\left(
\sigma ,\tau ,\varphi \right) h_{3[0]}\left( \sigma ,\tau \right) \delta
\varphi ^{2}  \label{sol4} \\
&&+\eta _{4}\left( \sigma ,\tau ,\varphi \right) h_{4[0]}\left( \sigma ,\tau
\right) \delta t^{2}\},  \nonumber
\end{eqnarray}
where (in order to construct a Class AA solution) we put
\begin{eqnarray*}
h_{3[0]} &=&a_{E}\left( \sigma ,\tau \right) a_{T}\left( \sigma ,\tau
\right) ,h_{4[0]}=b_{E}\left( \sigma ,\tau \right) b_{T}\left( \sigma ,\tau
\right) , \\
\eta _{4} &=&\omega ^{-2}\left( \sigma ,\tau ,\varphi \right) \varpi
^{-2}\left( \sigma ,\tau ,\varphi \right) ,
\end{eqnarray*}
considering anisotropic renormalizations of constants as in (\ref{oresc})
and (\ref{osect}). The polarization $\eta _{3}$ is to be found from the
relation
\begin{equation}
h_{3}=h_{[0]}^{2}[(\sqrt{|h_{4}|})^{\ast }]^{2},h_{[0]}^{2}=const,
\label{q1}
\end{equation}
which defines a soltuion of equation (\ref{ricci2a}) for $h_{4}^{\ast }\neq
0 $ , when $h_{3}=-\eta _{3}h_{3[0]}$ and $h_{4}=\eta _{4}h_{4[0]}.$
Substitutting the last values in (\ref{q1}) we get
\[
|\eta _{3}|=h_{[0]}^{2}\frac{b_{E}b_{T}}{a_{E}a_{T}}\left( \frac{\omega
^{\ast }+\varpi ^{\ast }}{\omega \varpi }\right) ^{2}.
\]
Then, computing the coefficient $\gamma ,$ see (\ref{abc}), after two
integrations on $\varphi $ we find
\begin{eqnarray*}
&& n_{i}\left( \sigma ,\tau ,\varphi \right) =n_{i[0]}\left(
\sigma ,\tau
\right) +n_{i[1]}\left( \sigma ,\tau \right) \int [\eta _{3}/\left( \sqrt{%
|\eta _{3}|}\right) ^{3}]d\varphi \\
&&=n_{i[0]}\left( \sigma ,\tau \right) +\tilde{n}_{i[1]}\left(
\sigma ,\tau \right) \int \omega \varpi \left( \omega ^{\ast
}+\varpi ^{\ast }\right) ^{2}d\varphi ,
\end{eqnarray*}
where we re-defined the function $n_{i[1]}\left( \sigma ,\tau
\right) $ into a new $\tilde{n}_{i[1]}\left( \sigma ,\tau \right)
$ by including all factors and constants like $h_{[0]}^{2},$
$b_{E},b_{T},a_{E}$ and $a_{T}.$

The constructed solution (\ref{sol4}) does not has as locally isotropic
limit the Schwarzschild metric. It has also a toroidal and ellipsoidal
horizons defined by the conditions of vanishing of $b_{E}$ and $b_{T},$ but
this solution is different from the metric (\ref{sol3}): it has a trivial
conformal factor and vanishing coefficients $\zeta _{i}$ which means that in
this case we are having a splitting of dynamics into three holonomic and one
anholonomic coordinate. We can select such functions $n_{i[0,1]}\left(
\sigma ,\tau \right) ,$ $\omega \left( \sigma ,\tau ,\varphi \right) $ and $%
\varpi \left( \sigma ,\tau ,\varphi \right) ,$ when at asymptotics one
obtains the Minkowski metric.

\section{Anisotropic Polarizations and Running Constants}

In this Section we consider non--static vacuum anho\-lo\-nomic
ellipsoidal and/or toroidal configurations de\-pen\-ding
explicitly on time variable $t$ and on holonomic coordiantes
$x^{i},$ but not on angular coordinate $\varphi .$ Such solutions
are generated by dynamical an\-ho\-lo\-nomic deformations and
conformal transforms of the Schwarzschild metric. For simplicity,
we analyze only Class A and AA solutions.

The coordinates are parametrized: $x^{i}$ are holonomic ones, in
particular, $x^{i}=\left( u,\lambda \right) ,$ for ellipsoidal
configurations, and $x^{i}=\left( \sigma ,\tau \right) ,$ for
toroidal configurations; $y^{3}=v=t$ and $y^{4}=\varphi .$ The
metric ansatz is stated in the form
\begin{eqnarray}
\delta s^{2} &=&\Omega ^{2}\left( x^{i},t\right) [-(dx^{1})^{2}-(dx^{2})^{2}
\nonumber \\
&&+h_{3}\left( x^{i},t\right) \delta t^{2}+h_{4}\left( x^{i},t\right) \delta
\varphi ^{2}],  \label{ansatz3}
\end{eqnarray}
where the differentials are elongated
\begin{eqnarray*}
d\varphi &\rightarrow &\delta \varphi =d\varphi +\zeta _{i}\left(
x^{k},t\right) dx^{i}, \\
dt &\rightarrow &\delta t=dt+n_{i}\left( x^{k},t\right) dx^{i}.
\end{eqnarray*}
The ansatz (\ref{ansatz3}) is related with some ellipsoidal and/ or toroidal
anholonomic deformations of the Schwarzschild metric (see respectively, (\ref
{schel}), (\ref{auxm1}), (\ref{auxm2}) and (\ref{schtor}), (\ref{auxm1t}), (%
\ref{auxm2t})) via time running renormalizations of ellipsoidal and toroidal
constants (instead of the static ones, (\ref{oresc}) and (\ref{osect})),
\begin{equation}
\rho _{g}\rightarrow \widehat{\rho }_{g}=\omega \left( x^{k},t\right) \rho
_{g},  \label{oresca}
\end{equation}
and
\begin{equation}
\rho _{g}^{[t]}\rightarrow \widehat{\rho }_{g}^{[t]}=\varpi \left(
x^{k},t\right) \rho _{g}^{[t]}.  \label{osecta}
\end{equation}
As particular cases we shall consider trivial values $\Omega ^{2}=1.$ The
horizons of such classes of solutions are defined by the condition of
vanishing of the coefficient $h_{3}\left( x^{i},t\right) .$

\subsection{Ellipsoidal/toroidal solutions with running constants}

\subsubsection{Trivial conformal factors, $\Omega ^{2}=1$}

The simplest way to generate a $t$--depending ellipsoidal (or
toroidal) configuration is to take the metric (\ref{auxm1}) (or
(\ref{auxm1t})) and to renormalize the constant as (\ref{oresca})
(or (\ref{osecta})). In result we obtain a metric (\ref{ansatz3})
with $\Omega ^{2}=1,$ $h_{3}=\eta _{3}\left( x^{i},t\right)
h_{3[0]}\left( x^{i}\right) $ and $h_{4}=h_{4[0]}\left(
x^{i}\right) ,$ where
\begin{eqnarray*}
\eta _{3} &=&\omega ^{-2}\left( u,\lambda ,t\right) ,h_{3[0]}=b_{E}\left(
u,\lambda \right) ,h_{4[0]}=a_{E}\left( u,\lambda \right) , \\
(\eta _{3} &=&\varpi ^{-2}\left( \sigma ,\tau ,t\right)
,h_{3[0]}=b_{T}\left( \sigma ,\tau \right) ,h_{4[0]}=a_{T}\left( \tau
\right) ).
\end{eqnarray*}
The equation (\ref{ricci2a}) is satisfied by these data because $h_{4}^{\ast
}=0$ and the condition (\ref{confeq}) holds for $\zeta _{i}=0.$ The
coefficient $\gamma $ from (\ref{abc}) is defined only by polarization $\eta
_{3},$ which allow us to write the integral of (\ref{ricci4a}) as
\[
n_{i}=n_{i[0]}\left( x^{i}\right) +n_{i[1]}\left( x^{i}\right) \int \eta
_{3}\left( x^{i},t\right) dt.
\]
The corresponding ellipsoidal (or toroidal) configuration may be transformed
into asymptotically Minkowschi metric if the functions $\omega ^{-2}\left(
u,\lambda ,t\right) $ (or $\varpi ^{-2}\left( \sigma ,\tau ,t\right) )$ and $%
n_{i[0,1]}\left( x^{i}\right) $ are such way determined by
boundary conditions that $\eta _{3}\rightarrow const$ and
$n_{i[0,1]}\left( x^{i}\right) \rightarrow 0,$ far away from the
horizons, which are defined by the conditions $b_{E}\left(
u,\lambda \right) =0$ (or $b_{T}\left( \sigma ,\tau \right) =0).$

Such vacuum gravitational configurations may be considered as to posses
running of gravitational constants in a local spacetime region. For
instance, in Ref \cite{v} we suggested the idea that a vacuum gravitational
soliton may renormalize effectively the constants, but at asymptotics we
have static configurations.

\subsubsection{Non--trivial conformal factors}

The previous configuration can not be related directly with the
Schwarzschild metric (we used \ its conformal transforms). A more direct
relation is possible if we consider non--trivial conformal factors. For
ellipsoidal (or toroidal) configurations we renromalize (as in (\ref{oresca}%
), or (\ref{osecta})) the conformal factor (\ref{auxm1c}) (or (\ref{auxm1tc}%
)),
\begin{eqnarray*}
\Omega ^{2}\left( x^{k},t\right) &=&\omega ^{2}\left( x^{k},t\right) \Omega
_{AE}^{2}(x^{k})b_{E}^{-1}\left( x^{k}\right) , \\
(\Omega ^{2}\left( x^{k},t\right) &=&\varpi ^{2}\left( x^{k},t\right) \Omega
_{AT}^{2}(x^{k})b_{T}^{-1}\left( x^{k}\right) ).
\end{eqnarray*}
In order to satisfy the condition (\ref{confq}) we choose $h_{3}=\Omega
^{-2} $ but $h_{4}=h_{4[0]}$ as in previous subsection: this solves the
equation (\ref{ricci2a}). The non--trivial values of $\zeta _{i}$ and $n_{i}$
are defined from (\ref{confeq}) and (\ref{ricci4a}),

\begin{eqnarray*}
\zeta _{i}\left( x^{k},t\right) &=&\left( \Omega ^{\ast }\right)
^{-1}\partial _{i}\Omega , \\
n_{i}\left( x^{k},t\right) &=&n_{i[0]}\left( x^{k}\right) +n_{i[1]}\left(
x^{k}\right) \int h_{3}\left( x^{i},t\right) dt.
\end{eqnarray*}
We note that the conformal factor $\Omega ^{2}$ is singular on horizon,
which is defined by the condition of vanishing of the coefficient $h_{3},$
i. e. of $b_{E}$ (or $b_{T}$). By a corresponding parametrization of
functions $\omega ^{2}\left( x^{k},t\right) $ (or $\varpi ^{2}\left(
x^{k},t\right) )$ and $n_{i[0,1]}\left( x^{k}\right) $ we may generate
asymptotically flat solutions, very similar to the Schwarzschild solution,
which have anholonomic running constants in a local region of spacetime.

\subsection{Black Elipsoid -- Torus Metrics with Running Constants}

Now we consider nonlinear superpositions of the previous metrics as to
construct solutions with running constants and two horizons (one ellipsoidal
and another toroidal).

\subsubsection{ Trivial conformal factor, $\Omega ^{2}=1$}

The simplest way to generate such metrics\ with two horizons is to
establish, for instance, a common toroidal system of coordinate,
to take the ellipsoidal and toroidal metrics constructed in
subsection V.A.1 and to multiply correspondingly their
coefficients. The corresponding data, defining a new solution for
the ansazt (\ref{ansatz3}), \ are
\begin{eqnarray}
g_{1,2} &=&-1,\widehat{\rho }_{g}=\omega \left( x^{k},t\right) \rho _{g},%
\widehat{\rho }
_{g}^{[t]}=\varpi \left( x^{k},t\right) \rho
_{g}^{[t]},\Omega =1,  \nonumber \\
h_{3} &=&\eta _{3}\left( x^{i},t\right) h_{3[0]}\left( x^{i}\right) ,\eta
_{3}=\omega ^{-2}\left( x^{k},t\right) \varpi ^{-2}\left( x^{k},t\right) ,
\nonumber \\
h_{3[0]} &=&b_{E}\left( x^{k}\right) b_{T}\left( x^{k}\right)
,h_{4}=h_{4[0]}=a_{E}\left( x^{i}\right) a_{T}(x^{i}),(%
  \nonumber \\
\zeta _{i} &=&0,n_{i}=n_{i[0]}\left( x^{k}\right) +n_{i[1]}\left(
x^{k}\right) \int \omega ^{-2}\varpi ^{-2}dt,  \label{data7}
\end{eqnarray}
where the  functions $a_{E},a_{T}$ and $b_{E}, b_{T}$ are given
by formulas  (\ref{auxm1s}) and (\ref{auxm1tb}).
 Analyzing the data (\ref{data7}) we conclude
that we have two horizons, when $b_{E}\left( x^{k}\right) =0$ and
$b_{T}\left( x^{k}\right) =0,$ parametrized respectively as
ellipsoidal and torus hypersurfaces. The boundary conditions on
running constants and on off--diagonal components of the metric
may be imposed as the solution would result in an asymptotic flat
metric. In a finite region of spacetime we may consider various
dependencies in time.

\subsubsection{Non--trivial conformal factor}

In a similar manner, we can multiply the conformal factors and coefficients
of the metrics from subsection V.A.2 in order to generate a solution
parametrized by the (\ref{ansatz3}) with nontrivial conformal factor $\Omega
$ and non-vanishing coefficients $\zeta _{i}.$ The data are
\begin{eqnarray}
g_{1,2} &=&-1,\widehat{\rho }_{g}=\omega \left( x^{k},t\right) \rho _{g},%
\widehat{\rho }_{g}^{[t]}=\varpi \left( x^{k},t\right) \rho _{g}^{[t]},
\label{data8} \\
\Omega ^{2} &=&\omega ^{2}\left( x^{k},t\right) \varpi ^{2}\left(
x^{k},t\right) \Omega _{AE}^{2}(x^{k})\times   \nonumber \\
&&\Omega _{AT}^{2}(x^{k})b_{E}^{-1}\left( x^{k}\right) b_{T}^{-1}\left(
x^{k}\right) ,(\mbox{ see (\ref{auxm1s}),(\ref{auxm1tb})}),  \nonumber \\
h_{3} &=&\Omega ^{-2},h_{3[0]}=b_{E}\left( x^{k}\right) b_{T}\left(
x^{k}\right) ,  \nonumber \\
h_{4} &=&h_{4[0]}=a_{E}\left( x^{i}\right) a_{T}(x^{i}),\zeta _{i}\left(
x^{k},t\right) =\left( \Omega ^{\ast }\right) ^{-1}\partial _{i}\Omega ,
\nonumber \\
n_{i} &=&n_{i[0]}\left( x^{k}\right) +n_{i[1]}\left( x^{k}\right) \int
\omega ^{-2}\varpi ^{-2}dt.  \nonumber
\end{eqnarray}
The data (\ref{data8}) define a new type of solution than that
given by (\ref {data7}). It this case there is a singular on
horizons conformal factor. The behaviour nearly horizons is very
complicated. By corresponding parametrizations of functions
$\omega \left( x^{k},t\right) ,$\ $\varpi \left( x^{k},t\right) $
and $n_{i[0,1]}\left( x^{k}\right) ,$ which approximate $\omega
,\varpi \rightarrow const$ and $\zeta _{i},n_{i}\rightarrow 0$ we
may obtain a stationary flat asymptotics.

Finally, we note that instead of Class AA solutions with anisotropic and
running constants we may generate solutions with two horizons (ellipsoidal
and toroidal) by considering nonlinear superpositions, anholonomic
deformations, conformal transforms and combinations of solutions of Classes
A, B. The method of construction is similar to that considered in this
Section.

\section{Conclusions and Discussion}

We constructed new classes of exact solutions of vacuum Einstein
equations by considering anholonomic deformations and conformal
transforms of the Schwarzshild black hole metric. The solutions
posses ellipsoidal and/ or toroidal horizons and symmetries and
could be with anisotropic renormalizations and running constants.
Some of such solutions define static configurations and have
Schwarzschild like (in general, multiplied to a conformal factor)
asymptotically flat limits. The new metrics are parametrized by
off--diagonal metrics which can be diagonalized with respect to
certain anholonomic frames. The coefficients of diagonalized
metrics are similar to the Schwazschild metric coefficients but
describe deformed horizons and contain additional dependencies on
one 'anholonomic' coordinate.

We consider that such vacuum gravitational configurations with non--trivial
topology and deformed horizons define a new class of ellipsoidal black hole
and black torus objects and/or their combinations.

Toroidal and ellipsoidal black hole solutions were constructed for different
models of extra dimension gravity and in the four dimensional (4D) gravity
with cosmological constant and specific configurations of matter \cite
{lemos,rev,emp}. There were defined also vacuum configurations for such
objects \cite{v,v2,vth,vtor}. However, we must solve the very important
problems of physical interpretation of solutions with anholonomy and to
state their compatibility with the black hole uniqueness theorems \cite
{israel} and the principle of topological censorship \cite{haw,cen}.

It is well known that the Scwarzschild metric is no longer the unique
asymptotically flat static solution if the 4D gravity is derived as an
effective theory from extra dimension like in recent Randall and Sundrum
theories (see basic results and references in \cite{rs}). The Newton law may
be modified at sub-millimeter scales and there are possible configurations
with violation of local Lorentz symmetry \cite{csaki}. Guided by modern
conjectures with extra dimension gravity and string/M--theory, we have to
answer the question:\ it is possible to give a physical meaning to the
solutions constructed in this paper only from a viewpoint of a generalized
effective 4D Einstein theory, or they also can be embedded into the
framework of general relativity theory?

It should be noted that the Schwarzschild solution was constructed as the
unique static solution with spherical symmetry which was connected to the
Newton spherical gravitational potential $\sim 1/r$ and defined as to result
in the Minkowski flat spacetime, at long distances. This potential describes
the static gravitational field of a point particle with ''isotropic'' mass $%
m_{0}.$ The spherical symmetry is imposed at the very beginning and it is
not a surprising fact that the spherical topology and spherical symmetry of
horizons are obtained for well defined states of matter with specific energy
conditions and in the vacuum limits. Here we note that the spherical
coordinates and systems of reference are holonomic ones and the considered
ansatz for the Schwarzschild metric is diagonal in the more ''natural''
spherical coordinate frame.

We can approach in a different manner the question of constructing 4D
static vacuum metrics. We might introduce into consideration off--diagonal
ansatz, prescribe instead of the spherical symmetry a deformed one
(ellipsoidal, toroidal, or their superposition) and try to check if a such
configurations may be defined by a metric as to satisfy the 4D vacuum Einstein
equations. Such metrics were difficult to be found because of cumbersome
calculus if dealing with off--diagonal ansatz. But the problem was
substantially simplified by an equivalent transferring of calculations with
respect to anholonomic frames \cite{v,vth,vtor}. Alternative exact static
solutions, with ellipsoidal and toroidal horizons (with possible extensions
for nonlinear polarizations and running constants), were constructed and
related to some anholonomic and conformal transforms of the Schwarzschild
metric.

It is not difficult to suit such solutions with the asymptotic limit to the
locally isotropic Minkowschi spacetime: ''an egg and/or a ring look like
spheres far away from their non--trivial horizons''. The unsolved question
is that what type of modified Newton potentials should be considered in this
case as they would be compatible with non--spherical symmetries of
solutions? The answer may be that at short distances the masses and
constants are renormalized by specific nonlinear vacuum gravitational
interactions which can induce anisotropic effective masses, ellipsoidal or
toroidal polarizations and running constants. For instance, the Laplace
equation for the Newton potential can be solved in ellipsoidal coordinates
\cite{ll}: this solution could be a background for constructing ellipsoidal
Schwarzshild like metrics. Such nonlinear effects should be treated, in some
approaches, as certain quasi--classical approximations for some 4D quantum
gravity models, or related to another type of theories of extra dimension
classical or quantum gravity.

Independently of the type of little, or more, internal structure of black
holes with non--spherical horizons we search for physical justification, it
is a fact that exact vacuum solutions with prescribed non--spherical
symmetry of horizons can be constructed even in the framework of general
relativity theory. Such solutions are parametrized by off--diagonal metrics,
described equivalently, in a more simplified form, with respect to
associated anholonomic frames; they define some anholonomic vacuum
gravitational configurations of corresponding symmetry and topology.
Considering certain characteristic initial value problems we can select
solutions which at asymptotics result in the Minkowschi flat spacetime, or
into an anti--de Sitter (AdS) spacetime, and have a causal behaviour of
geodesics with the equations solved with respect to anholonomic frames.

It is known that the topological censorship principle was reconsidered for
AdS black holes \cite{cen}. But such principles and uniqueness black hole
theorems have not yet been proven for spacetimes defined by generic
off--diagonal metrics with prescribed non--spherical symmetries and horizons
and with associated anholonomic frames with mixtures of holonomic and
anholonomic variables. It is clear that we do not violate the conditions of
such theorems for those solutions which are locally anisotropic and with
nontrivial topology in a finite region of spacetime and have locally
isotropic flat and trivial topology limits. We can select for physical
considerations only the solutions which satisfy the conditions of the
mentioned restrictive theorems and principles but with respect to well
defined anholonomic frames with holonomic limits. As to more sophisticate
nonlinear vacuum gravitational configurations with global non--trivial
topology we conclude that there are required a more deep analysis and new
physical interpretations.

The off--diagonal metrics and associated anholonomic frames extend the class
of vacuum gravitational configurations as to be described by a nonlinear,
anholonomic and anisotropic dynamics which, in general, may not have any
well known locally isotropic and holonomic limits. The formulation and proof
of some uniqueness theorems and principles of topological censorship as well
analysis of physical consequences of such anholonomic vacuum solutions is
very difficult. We expect that it is possible to reconsider the statements
of the Israel--Carter--Robinson and Hawking theorems with respect to
anholonomic frames and spacetimes with non--spherical topology and
anholonomically deformed spherical symmetries. These subjects are currently
under our investigation.

\subsection*{Acknowledgements}

\begin{acknowledgments}
The work is partially supported by a
2000--2001 California State University Legislative Award and by a
NATO/Portugal fellowship grant at the Technical University of
Lisbon. The author is grateful to J. P. S. Lemos,  D. Singleton
and E. Gaburov for support and collaboration.
\end{acknowledgments}


\end{document}